\documentclass[manuscript]{emulateapj}
\usepackage{graphicx}

\slugcomment{Accepted by AJ July 4, 2009}

\begin{document}

\title{Faint, Evolving Radio AGN in SDSS Luminous Red Galaxies}

\shorttitle{Faint, Evolving Radio AGN in SDSS LRGs}

\author{J. A. Hodge, G. R. Zeimann, R. H. Becker\altaffilmark{1}}
\affil{University of California, 1 Shields Ave, Davis, CA 95616}
\email{hodge@physics.ucdavis.edu}

\author{R. L. White}
\affil{Space Telescope Science Institute, 3700 San Martin Drive, Baltimore, MD 21218}

\altaffiltext{1}{Lawrence Livermore National Laboratory, L-413, Livermore, CA 94550}

\begin{abstract}
We detect and study the properties of faint radio AGN in Luminous Red Galaxies (LRGs).  The LRG sample comprises 760,000 objects from a catalog of LRG photometric redshifts constructed from the Sloan Digital Sky Survey (SDSS) imaging data, and 65,000 LRGs from the SDSS spectroscopic sample.  These galaxies have typical 1.4 GHz flux densities in the 10s-100s of $\mu$Jy, with the contribution from a low-luminosity AGN dominating any contribution from star formation.  To probe the radio properties of such faint objects, we employ a stacking technique whereby FIRST survey image cutouts at each optical LRG position are sorted by the parameter of interest and median-combined within bins.  We find that median radio luminosity scales with optical luminosity ($L_{opt}$) as $L_{1.4 GHz} \propto L_{opt}^{\beta}$, where $\beta$ depends on the redshift being probed.  Above $z \simeq 0.4$, $\beta$ appears to decrease from $\beta \simeq 1$ at $z = 0.4$ to $\beta \simeq 0$ at $z = 0.7$, a result which could be indicative of AGN cosmic downsizing.  We also find that the overall LRG population, which is dominated by low-luminosity AGN, experiences significant cosmic evolution between $z = 0.2$ and $z = 0.7$.  A simultaneous fit to untangle the redshift and luminosity dependencies yields redshift evolution of the form $L_{1.4GHz} \propto (1+z)^{3.15 \pm 0.07}$, implying a considerable increase in total AGN heating for these massive ellipticals with redshift.  By matching against the FIRST catalog, we investigate the incidence and properties of LRGs associated with double-lobed (FR I/II) radio galaxies.  

 

\end{abstract}

\section{INTRODUCTION}
\label{Intro}

Luminous Red Galaxies (LRGs) are among the most luminous galaxies in the universe, with L $\ge$ 3L$^{\ast}$ \citep{eis01}.  These massive ellipticals have passively-evolving stellar populations and remarkably uniform SEDs characterized by a strong 4000-\AA\ break.  It is the redshifting of this break through the different filters that gives LRGs their red colors and enables their discovery via photometry over a large redshift range.  

Little is currently known about the radio properties of these massive ellipticals.  The vast majority of LRGs are not powerful radio galaxies, but their nuclei are by no means uniformly radio-silent. The supermassive black holes hosted by the LRG population are likely remnants of more powerful active galactic nuclei (AGN) that formed at an earlier epoch and now exist as low-luminosity radio AGN.  While high-power radio sources show strong evolution (e.g. Willott et al. 2001), the evolution of low-luminosity radio AGN ($L_{1.4GHz}$ $<$ 10$^{33}$ ergs s$^{-1}$ Hz$^{-1}$) is still a somewhat controversial topic  (Sadler et al. 2007; Cowie et al. 2004; Smolcic et al. 2009).  Setting limits on the radio emission from these LRGs can therefore tell us something about the evolution of low-luminosity radio AGN.  

As the evolution of galaxies is thought to be tied to the evolution of their central black holes, this study also has implications for galaxy formation/evolution.  The radio is particularly useful here as it is not affected by dust obscuration (although orientation can be an issue).  In addition, `radio-mode' AGN feedback has recently gained fame as an important addition to semi-analytic models of galaxy formation and evolution, with the mechanical energy input from the central radio source effectively suppressing star formation in massive galaxies at late times and explaining the exponential cutoff at the bright end of the galaxy luminosity function (Croton et al. 2006; Cattaneo \& Teyssier 2007).  Currently, no clear picture exists for how this radio mode works, but evidence is mounting that the mechanical energy injected into the surrounding medium by radio sources may balance the radiative energy losses from the hot gas (Best et al. 2006; Smolcic et al. 2009).  In massive red galaxies such as these, then, one would expect to find radio AGN to explain the lack of ongoing star-formation.  Quantifying the level of radio emission and understanding the evolution of this population with redshift are therefore important for understanding the role of radio AGN feedback in massive red galaxies.  

Accomplishing this task is not entirely straighforward.  The vast majority of LRGs are radio-quiet and lie far below the detection threshold of wide-area surveys like the FIRST survey.  A significant portion would remain undetected even in current `deep' surveys.  Therefore, to study the entire LRG population in the radio, we will use a median-stacking technique on LRGs optically-selected from the Sloan Digital Sky Survey (SDSS).  Median-stacking has been used previously to investigate the radio emission of various classes of optically-selected objects.  The technique we use was first introduced by \citet{whi07}, hereafter Paper I, where they quantified the systematic effects associated with stacking FIRST images and examined the median radio properties of quasars from the SDSS.  \citet{deV07}, hereafter Paper II, then used the same method to study low-luminosity AGN.  Finally, in \citet{hod08}, hereafter Paper III, the technique was applied to look at SDSS galaxies lacking strong emission lines (optically-`quiescent' galaxies).  Over $\sim$60,000 of these galaxies were part of the spectroscopic LRG sample.  (Indeed, the vast majority of LRGs are optically-`normal'.)  By median-stacking radio maps centered on these quiescent LRGs, Paper III found evidence for low-level AGN activity in the form of sub-mJy radio emission. We will use this finding, along with a catalog of LRG photometric redshifts almost ten times as large, to study the dependence of radio properties on the mass and redshift of the host galaxy.

While LRGs tend to harbor low-luminosity radio AGN, a small percentage are radio-loud.  It has been known for some time that powerful radio galaxies are associated with massive ellipticals, and the probability that an early-type galaxy will be a radio galaxy increases with optical luminosity (e.g., Matthews, Morgan, \& Schmidt 1964;  Auriemma et al. 1977).  Still, it is not well understood what makes particular galaxies radio-loud while others are not, or what factors cause certain galaxies to exhibit the distinctive double-lobed radio structure of FR I and FR II galaxies.  By matching the optical LRG positions to their complete radio environment in FIRST, we will study the occurrence of double-lobed radio galaxies in the LRG population.  


We begin in \S \ref{Data} by going over the sample compilation (\S \ref{sample}) and radio data (\S \ref{data}).   When then discuss our analysis, including our radio stacking technique (\S \ref{stacking}) and radio galaxy matching (\S \ref{matching}).  Our results are presented in \S \ref{results}, with median radio properties of the entire LRG sample discussed in \S \ref{median}, and the matching to double-lobed radio galaxies discussed in \S \ref{FR2}.   A more detailed discussion follows in \S \ref{Discussion}, and we end with our conclusions in \S \ref{conclusions}.  Where applicable we assume the standard FRW cosmology of H$_0$ = 70 km s$^{-1}$ Mpc$^{-1}$, $\Omega_{\Lambda}$ = 0.7, and $\Omega_{M}$ = 0.3.







\section{THE DATA}
\label{Data}
\subsection{Sample Compilation}
\label{sample}

The main LRG sample comes from a preliminary version of a catalog of LRG photometric redshifts from the SDSS \citep{yor00}.  The catalog construction is discussed in \citet{pad05}.  As opposed to the MegaZ-LRG catalog \citep{coll07} which uses an artificial neural network to compute photometric redshifts, this catalog uses a template based method.  This enables more robust extrapolation beyond the training set.  We are using the most recent version of the catalog, which currently contains approximately 1.1 million LRGs, of which 764,870 have overlapping FIRST coverage (see \S \ref{data} below).  The photometric redshifts span the range $0.1 < z < 0.99$.  For redshifts up to 0.6, these values have a scatter of $\sim$3\% and biases of up to a few percent due to photometric errors and systematic uncertainties in the templates \citep{pad05}.  For $0.6 < z < 0.7$, the scatter increases to $\sim$5\%, and it worsens considerably above $z = 0.7$.  The selection criteria used for this version of the catalog are those of the 2dF-SDSS LRG and QSO survey, 2SLAQ \citep{can06}.  We will refer to this sample as the ``photo-z LRG" sample.

We also use objects targeted as LRGs in the SDSS spectroscopic survey \citep{eis01}.  This sample comes from the NYU Value-Added Galaxy Catalog (NYU-VAGC), a collection of galaxy catalogs derived from the SDSS DR4 \citep{bla05}.  As suggested in \citet{eis01}, we keep only those objects with the GALAXY\_RED flag set and redshifts $z > 0.2$, restrictions which help account for the breakdown of the luminosity cut at lower redshifts.  This sample is less than 10\% the size of the photo-z LRG sample, and we will refer to it as the ``spec-z LRG" sample.

We use SDSS model magnitudes throughout this paper.  To calculate extinction-corrected absolute magnitudes for the photo-z LRG sample, we use version v4.1.4 of the IDL kcorrect code from \citet{bla03}.  For the spec-z LRG sample, we calculate magnitudes from the extinction-corrected fluxes available in the kcorrect catalog associated with the NYU-VAGC \citep{bla07}.  As the combined sample spans a large enough redshift range to make evolutionary changes a concern, we then generate k$+$e corrections simultaneously using the \citet{bru03} stellar population synthesis code.  In the optical, \citet{wak06} found no evidence for any additional evolution of the luminosity function of LRGs beyond that expected from passive evolution of the stellar population.  We therefore assume a single instantaneous burst of star formation at $z = 9.84$ of solar metallically and passively evolve the stellar population, as done in \citet{ros06}.  This simple model has been found to adequately approximate the color evolution of LRGs \citep{wak06}.   

\begin{figure}[ht]
\centering
\includegraphics[scale=0.65]{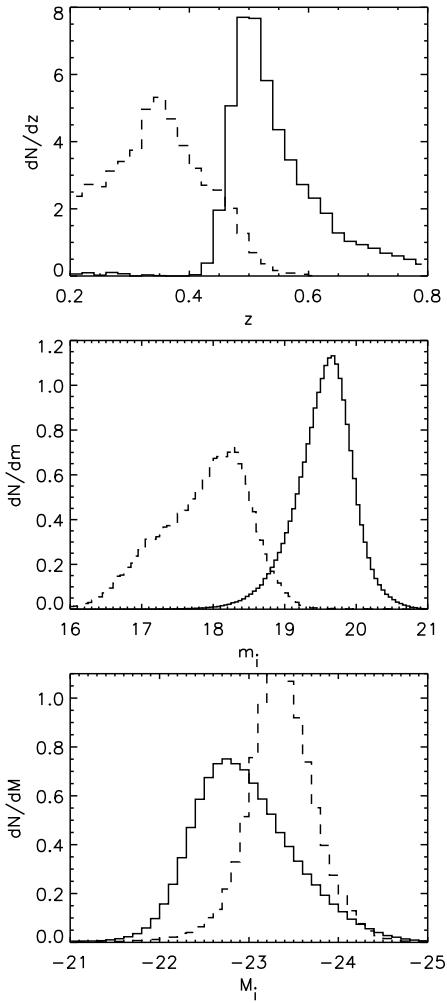}
\caption{Overview of the properties of the photo-z (solid) and spec-z (dashed) LRG samples.  The histograms show (from top): redshift, extinction-corrected apparent magnitude in the i-band, and absolute magnitude in the i-band (k$+$e corrected).  All distributions are normalized to integrate to unity.}
\label{fig:properties}
\end{figure}

Figure \ref{fig:properties} gives a summary of the basic properties of the LRG samples.  The distributions have all been normalized such that they integrate to unity.  As Figure \ref{fig:properties} shows, the galaxies in the photo-z LRG sample are intermediate redshift and typically fainter than the galaxies in the SDSS spec-z LRG sample.  The redshift histogram shows a steep rise at $z = 0.4$ for the photo-z LRG sample, highlighting the effectiveness of the 2SLAQ selection in targeting intermediate-redshift LRGs.  The absolute magnitudes of the photo-z and spec-z LRG samples have been k$+$e-corrected to $z = 0.0$, and the two samples cover an overlapping range in absolute magnitude.  



\subsection{Radio Data}
\label{data}

The radio data that we use for the median-stacking come from the VLA FIRST survey \citep{beck95}.  The FIRST survey is a 1.4 GHz survey of 9,055 $\deg^2$ of the North Galactic Cap and Equatorial Strip.  With pixels of 1.8", the survey has a resolution of 5".  The FIRST data were obtained in 160 sec VLA snapshots and have a typical rms of 0.15 mJy beam$^{-1}$.  To convert to rest-frame 1.4 GHz luminosity for each galaxy, we perform a k-correction assuming an average radio spectral index $\alpha = -0.5$ as in Papers II and III.  When calculating luminosities for our sample of double-lobed radio galaxies, we use a steeper index of -0.75, as the lobes dominate the flux density.  This is consistent with that used in \citet{deV06}.  For our radio-matching analysis, we use the newest version of the FIRST catalog (08Jul16 Version), which includes information on counterparts in the SDSS.  

\section{ANALYSIS}
\subsection{Median Stacking}
\label{stacking}

The FIRST survey is not deep enough to see all, or even most, of these LRGs individually.   We therefore employ the median-stacking technique first presented in Paper I to ``see below the noise".  Taking our optically-selected sample, we retrieve a square FIRST cutout centered on the position of each LRG.  The cutouts may either be left in units of flux density, or, if redshifts are known, converted to units of luminosity prior to stacking (Paper I).  Depending on the number of cutouts that are stacked, we can achieve various levels of sensitivity.  Stacking 3000 cutouts produces an image with a theoretical rms of 3 $\mu$Jy beam$^{-1}$, a number that is comparable with the deepest current surveys.  With a large enough sample, such as we have here, we can sort the cutouts by some parameter of interest (for e.g., absolute magnitude of the galaxy) and median-stack the cutouts within each bin separately.  Note that we are stacking \textit{different} galaxies together.  By binning the sorting parameter in the same manner and calculating the median value per bin, we can examine general trends in radio flux density (or luminosity) of very faint objects as a function of said parameter.  

For unresolved sources, the radio luminosity is equivalent to the peak of the emission.  As in Papers I through III, then, we derive flux densities from the value of the central pixel, and thus the size of each cutout is of little importance.  We corrected for the `snapshot bias' present for sub-threshold sources (Paper I) by multiplying by 1.40.  To compute error bars, we used binomial probabilities as described in \citet{gott01}.  For more on the use of medians and the procedure in general, see Paper I. 

A noteworthy comment about stacking in this way is that the stacked radio data are not flux limited.  As cutouts around every optical source are included in the calculation of the median, there is no intrinsic bias against faint radio sources despite the fact that FIRST is itself flux limited.  Any observed bias must be attributed to the optical flux limit passed down second-hand through a radio-optical correlation.  

\begin{figure}[ht]
\centering
\includegraphics[scale=0.5]{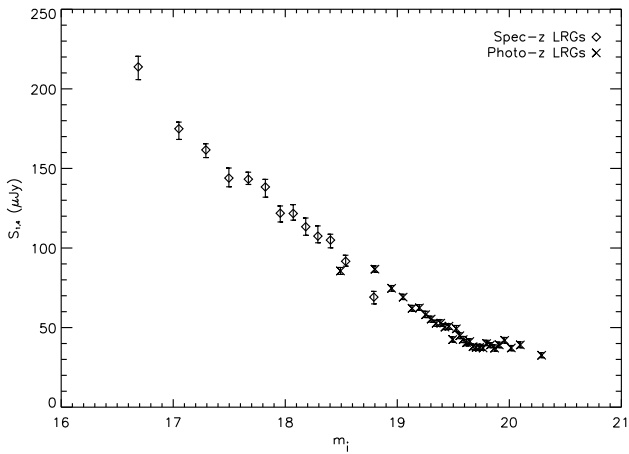}
\caption{Median stacked 1.4 GHz flux density versus m$_i$.  A bin size of 5,000 galaxies was used for the smaller spec-z LRG sample, while a bin size of 25,000 was used for the photo-z LRG sample to minimize cluttering.}
\label{fig:flux}
\end{figure}

Figure \ref{fig:flux} shows median flux density versus m$_i$ for the two LRG samples. It resulted from sorting/stacking each sample by i-band apparent magnitude following the above method and using a bin size of 5,000 galaxies for the spec-z LRG sample and 25,000 for the much larger photo-z LRG sample (to reduce clutter).  The photo-z LRG sample appears to make up the faint-end tail of the spec-z LRG sample.  It is also apparent from this figure that the majority of these galaxies fall far below the FIRST catalog detection threshold of 1 mJy, but in every bin a clear detection is achieved.  Only around 10\% of the spec-z LRG sample have catalogued FIRST sources within 3", while fewer than 3\% of the photo-z LRG sample have matches. 


\subsection{Radio Galaxy Matching}
\label{matching}

In order to study the occurrence of double-lobed radio galaxies in the LRG population, we took the LRG sample and matched them with the surrounding FIRST radio environment to look for pairs of matching radio sources.  As random radio sources may sometimes fall within the search radius, masquerading as lobes, we have taken several precautions to maximize the reliability of matches.  As a zeroth-order safety measure, we have first discarded any potential lobes with optical counterparts from SDSS within 1.5", as true radio lobes would have no optical signature.  We have then utilized the weighting scheme of \citet{deV06} to rank the possible configurations by their likelihood of being a true radio lobe.  Nearby, unrelated sources tend to have small opening angles, so this procedure weights against opening angles smaller than $50^{\circ}$.    We did not require the core to have a match (defined as within 3"), as this would bias against LRGs without detectable cores.  

For the initial match, we use a search radius of 450".  As the incidence of false matches increases with increasing search radius, the percentages we quote throughout this paper include only FR I/II candidates with lobe separations less than 350"  (see the discussion in Section \ref{FR2}).  As a control, we repeat the matching procedure on a list of random positions, created by adding 0.5$^{\circ}$ to the LRG positions.  We do this in eight different directions and average over the results.  We take the excess of matches around LRG positions over matches around control positions as an estimate of the true number of double-lobed radio galaxy matches.

\begin{figure*}[ht]
\centering
\includegraphics[scale=0.8]{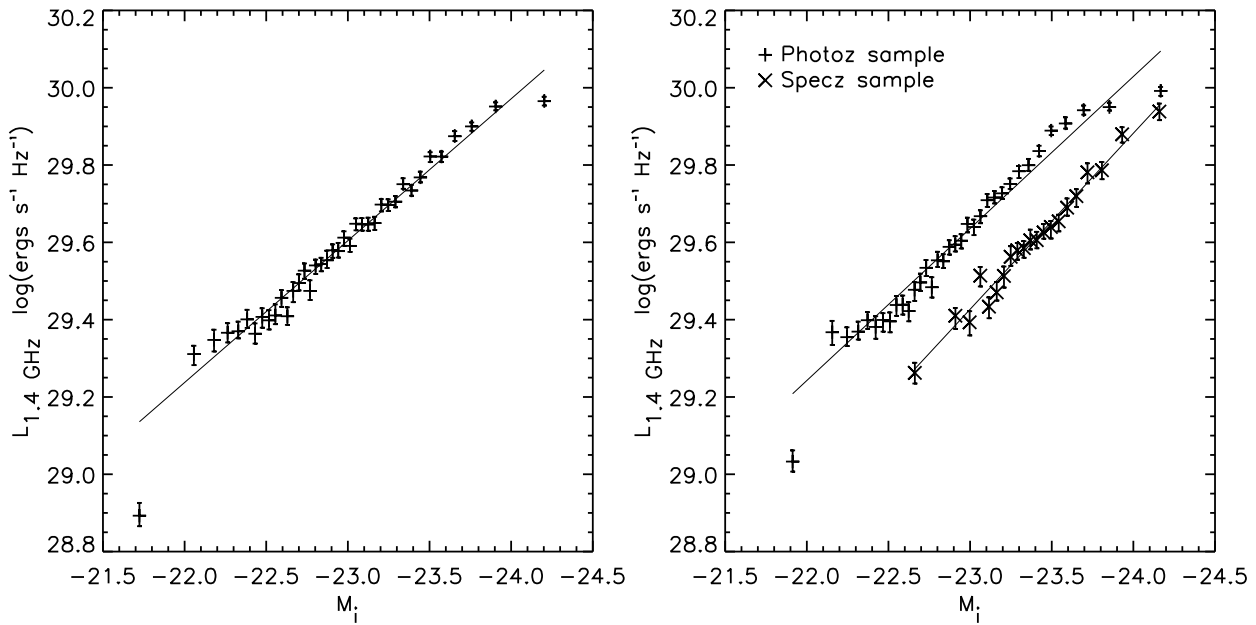}
\caption{Rest-frame 1.4 GHz radio luminosity vs. i-band absolute magnitude for all LRGs.  Bins are 20,000 LRGs each.  The black line shows the best fit to the data.}
\label{fig:rlumvsMi_all}
\end{figure*}

\section{RESULTS}
\label{results}

\subsection{Median Radio Properties}
\label{median}

As discussed in the Introduction, Paper III used median stacking to detect low level radio emission from a sample of spectroscopically-targetted LRGs that were classified as quiescent on the basis of optical emission line ratios \citep{brinch04}.  (As noted in the Introduction, this describes the vast majority of LRGs.)  This work also provided evidence that the faint radio emission detected in these LRGs is the result of AGN activity, giving us a first glimpse at the average level of activity in the sample as a whole.  Here we continue the study of LRGs, this time with a much larger sample extending up to intermediate redshifts.

\subsubsection{A Note on Using Median Radio Luminosity}
\label{duty cycle}

If the radio emission from LRGs is indeed due to low-luminosity AGN activity, then the question becomes how to interpret the resulting values of median radio luminosity.  Massive galaxies are thought to cycle between radio-loud and radio-quiet phases caused by hot intergalactic gas cooling to fuel the central AGN, then being reheated by the subsequent radio jets.  While we have no prior knowledge of the length of a typical duty cycle, the fraction of galaxies above some radio luminosity can be thought of as the fraction of the duty cycle for which a particular galaxy is emitting at or above this level.  A time-average can therefore be substituted for an ensemble-average, and the median radio luminosity can be thought of as the average radio luminosity of a representative LRG.

When interpreting a change in median luminosity for the sample, things get a bit trickier.  The existence of two distinct types of objects in the sample (radio AGN ``on" or ``off") means that a change in the median luminosity has a range of possible physical causes, from a change in the luminosities of the sources, to a change only in the length of the duty cycles.  This should be kept in mind when interpreting the results that follow.

\subsubsection{Radio-Optical Correlation}
\label{downsizing}

As mentioned in Section \ref{stacking}, a radio-optical correlation could cause a bias against faint radio sources.  We therefore wish to first see how radio luminosity depends on optical luminosity for the LRGs.  To accomplish this, we sort and bin the data by i-band absolute magnitude.  We will use i-band magnitude here as the i-band filter lies above the 4000-\AA\ break for our redshift range and thus the K-corrections are less sensitive to redshift.  We plot median radio luminosity versus $M_i$ in Figure \ref{fig:rlumvsMi_all} (left panel).  The bin size in this figure has been set at 20,000 galaxies to avoid excessive crowding of the data points, and we have cut LRGs with $z > 0.7$ due to the increased uncertainties in the photometric redshift estimation.  Although the figure shows deviation from a power law at high and low luminosities, these points come from the tails of the distribution where outliers are more likely to dominate.  It thus appears that, for the most part, the median radio luminosity increases with increasing median absolute magnitude.  When we translate i-band magnitude into luminosity, we find that $L_{radio} \propto L_{optical}^{\beta}$ with $\beta = 0.92 \pm 0.01$.



In the right panel of Figure \ref{fig:rlumvsMi_all}, we show the photo-z and spec-z LRG samples separately.  The photo-z LRG sample is still grouped in bins of 20,000, while the smaller spec-z LRG sample has a bin size of 3,000.  Here we get the first hint that the radio luminosity of LRGs evolves with redshift, as the two LRG samples have different mean redshifts (0.35 for the spec-z sample versus 0.55 for the photo-z sample).  The $\Delta$log(L$_{1.4GHz}$) = 0.2 difference implies radio evolution of the form L$_{1.4GHz}$ $\propto$ (1+z)$^{10/3}$.   We will investigate the form of the evolution further in Section 4.1.3.  The dependence on optical luminosity for the two samples are $\beta = 0.98 \pm 0.01$ for the photo-z LRG sample and $\beta = 1.14 \pm 0.04$ for the spec-z LRG sample.  Note that the slope of the combined sample is shallower than either sample individually.  This is likely due to the fact that photo-z LRGs dominate above M$_i$ = -22.5, pulling that end up and requiring a shallower fit.

\begin{figure*}[ht]
\centering
\includegraphics[scale=0.8]{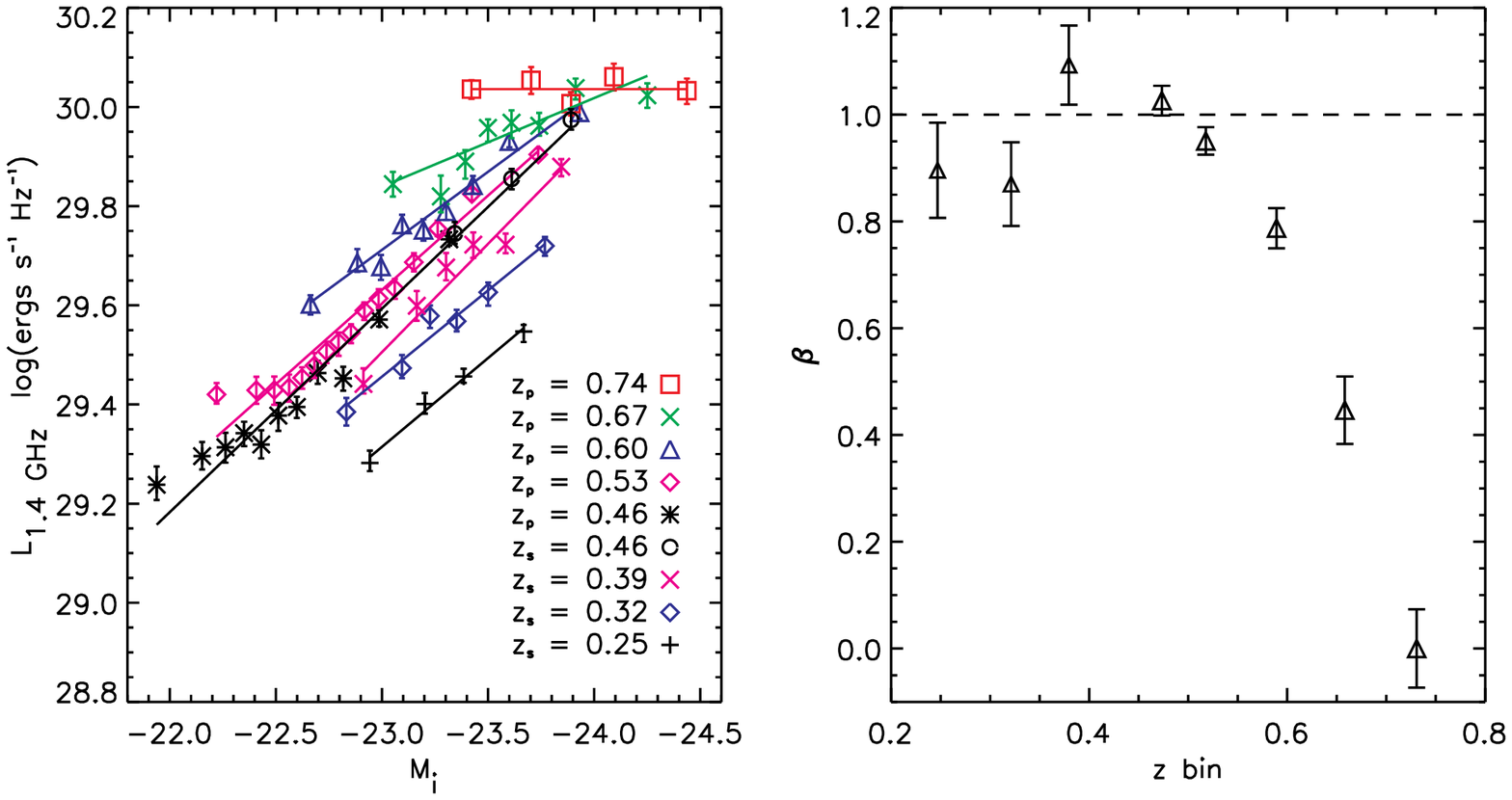}
\caption{Left panel:  Rest-frame 1.4 GHz radio luminosity vs. i-band absolute magnitude for LRGs grouped by redshift (z$_p$ = photo-z, z$_s$ = spec-z).  The bin size has been adjusted as necessary to reduce clutter in the plot, and best-fit lines are overplotted.  Right panel:  $\beta$ vs. median redshift for each of the redshift bins, where $L_{radio} \propto {L_{optical}}^{\beta}$.  The dashed line shows the value of $\beta$ when all the data are fit together, as in Figure \ref{fig:rlumvsMi_all}.}
\label{fig:rlumvsMi_slopes}
\end{figure*}

As absolute magnitude is strongly correlated with redshift, we have next divided the LRGs into narrow redshift groups within the spec-z and photo-z samples.  The results are shown in Figure \ref{fig:rlumvsMi_slopes}.  Here we have left in the $z > 0.7$ galaxies since they are in a separate group and thus do not contaminate the remaining data (as long as the reader keeps in mind that the $z = 0.74$ group may not be reliable).  The left panel shows the correlations for individual redshift groups (where the bin size has been adjusted as necessary for clarity), while the right panel shows the subsequent $\beta$ values assuming $L_{radio} \propto L_{optical}^{\beta}$ as above.  We have overlaid a dashed line representing $L_{radio} \sim L_{optical}$, close to what was derived for the (significantly larger) photo-z LRG sample.  Note that both the photo-z and spec-z LRG samples cover the redshift group $z = 0.46$, and we have kept the samples separate as a consistency check.  While the spec-z LRG sample covers a slightly brighter range in M$_i$, we take the excellent agreement at M$_i$ = -23.3 (there are two overlapping data points there) as assurance that the two samples are comparable.  The $z = 0.46$ data for both samples has been fit with one line.  

\begin{figure*}[ht]
\centering
\includegraphics[scale=0.75]{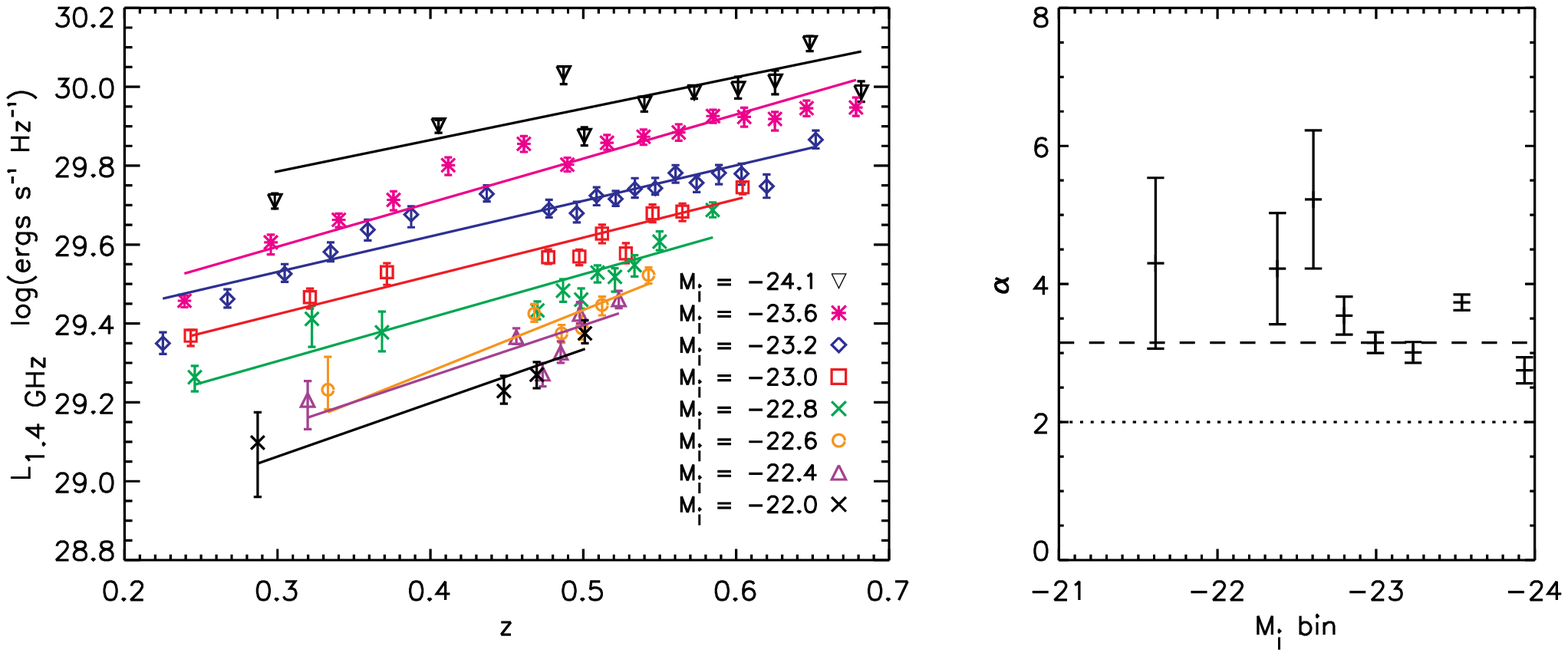}
\caption{Left panel:  Rest-frame 1.4 GHz radio luminosity vs. redshift for LRGs grouped by i-band absolute magnitude.  The bin size has been adjusted as necessary to reduce clutter in the plot, and best-fit lines are overplotted.  The redshift range was restricted to $z < 0.7$ to minimize errors in photometric redshift.  Right panel:  Values of $\alpha$ vs. median absolute magnitude for the magnitude bins shown in the left panel, where $L_{radio} \propto (1+z)^{\alpha}$.  The dotted line is the luminosity evolution derived by \citet{sad07} for LRGs associated with low-luminosity radio galaxies detected in FIRST.  The dashed line is our result from simultaneously fitting the redshift and optical luminosity dependencies.}
\label{fig:rlumvsz}
\end{figure*}

Figure \ref{fig:rlumvsMi_slopes} implies that slope of $\beta = 1$ seen for the combined data was likely due to the fact that lower-redshift galaxies dominate the sample (binning is variable in this figure, but this assertion remains true.)  We also wish to point out the large error bars on the redshift bins with purely spectroscopic data.  These points are the three points left of $z = 0.4$ in the right panel, and the large uncertainty is due to the small ranges in M$_i$ covered.  The points rightward of $z = 0.4$ appear to demonstrate that optically-fainter AGNs exhibit stronger evolution in radio luminosity than optically-brighter AGNs.  We will return to this point in Section \ref{Discussion}.

\subsubsection{Redshift Dependence}

We next look to see if the radio luminosity of the LRGs evolves with redshift.  The left panel of Figure \ref{fig:rlumvsz} shows median radio luminosity plotted as a function of redshift, where we have cut LRGs with $z > 0.7$ to minimize scatter due to errors in the photometric redshift estimates.  The absolute magnitude groups are defined in the legend.  This plot shows that the median radio luminosity for a given absolute magnitude bin increases with redshift, indicating that the LRGs are experiencing cosmic evolution of radio power over the redshift range $0.2 < z < 0.7$.  Fitting $L_{radio} \propto (1+z)^\alpha$ gives values of $\alpha$ ranging from around 3 to 5 (with error bars: see the right panel of Figure \ref{fig:rlumvsz}), with a mean of $\alpha = 3.7 \pm 0.3$.  

One might worry that the strength of this evolution is being artificially inflated by the radio-optical correlation found in the previous section.  Despite the fact that we have attempted to minimize this bias by grouping the sample by absolute magnitude, the finite width of the absolute magnitude group means that data points within a group may still have a range in median $M_i$ values, which would then be affected by the correlation found between optical and radio luminosity.  However, due to the use of two different samples (photo-z and spec-z LRGs), the median $M_i$ values for bins within an $M_i$ group do not rise monotonically with increasing redshift, but rather, the spec-z LRG bins tend to have slightly higher values of median $M_i$ than the photo-z LRG bins.  Nevertheless, the median $M_i$ \textit{within} the photo-z LRG sample, and \textit{within} the spec-z LRG sample (separately) does increase with redshift, and so we have restricted the dispersion within an absolute magnitude group to 0.06 mags or less.  Based on this dispersion in $M_i$, and applying the relation we derived above for the radio-optical correlation of the sample altogether, we estimate that the intrinsic radio-optical correlation makes up $\le$ 10\% of the increase in radio luminosity in any given absolute magnitude group.  

Although the effect of the radio-optical correlation is largely removed by the technique of grouping by absolute magnitude, the $\le$ 10\% effect remaining could still be significantly affecting the value of $\alpha$ measured.  We therefore apply the next simplest model to untangle the various dependencies.  We assume a simple power-law dependence on both optical luminosity and redshift, with no cross-terms:
\begin{equation}
\mathrm{log}(L_{1.4GHz}) = A + \beta \ \mathrm{log}(L_{opt}) + \alpha \ \mathrm{log}(1+z) 
\end{equation}
We bin the data by $L_{opt}$ and z on a two-dimensional irregular grid and perform a least-squares fit.  We again limit the redshift range to z $<$ 0.7 since those values have large uncertainties.  The resulting fit is less than ideal, with a reduced $\chi^2$ of 4.7, implying that our simplistic model might be too simple.  The best-fit parameters are $\beta = 0.89 \pm 0.01$ and $\alpha = 3.15 \pm 0.07$.  The dependence on optical luminosity is therefore consistent, within error bars, with that derived from Figure \ref{fig:rlumvsMi_all}, leading us to conclude that the redshift binning in that figure did a fairly good job removing the dependence on redshift evolution.  The value of $\alpha$ we determine is also consistent, within error bars, with the mean value we obtained from the fits in Figure \ref{fig:rlumvsz}, but the uncertainty has been cut in half.  This $\alpha$ value still implies significant redshift evolution, and we overplot this result in Figure \ref{fig:rlumvsMi_slopes} as a dashed line.  

As discussed in Section \ref{duty cycle}, the use of median radio luminosity masks the physical mechanism responsible for the evolution we measure.  Due to the finite duty cycle of AGN activity, there 
is more than one way to achieve an increase in median radio luminosity.  The evolution seen here could have resulted from an increase in average luminosity and/or an increase in average duty cycle length.  The two extremes (all sources get brighter with the same duty cycles, or the duty cycles increase for unchanging luminosities) correspond to pure luminosity and pure density evolution respectively, and we cannot distinguish between them using this method.

\subsection{Double-lobed Radio Galaxies}
\label{FR2}

\subsubsection{Searching for FR I/IIs}

\begin{figure}[ht]
\centering
\includegraphics[scale=0.43]{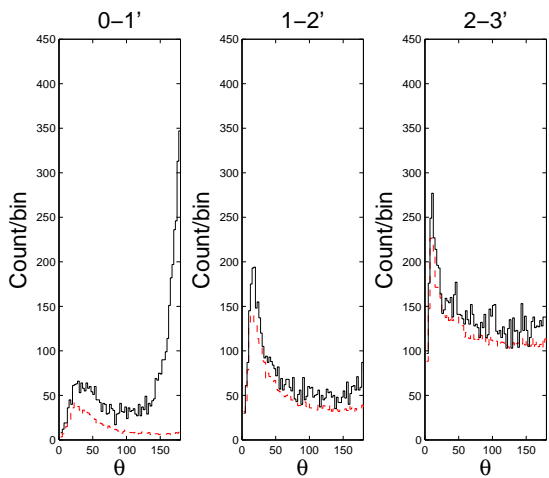}
\caption{Number of radio galaxies versus lobe opening angle ($\theta$) for the photo-z LRG sample, with the three panels showing lobe separations of 0-1', 1-2', and 2-3'.  The black (solid) and red (dashed) lines show the counts/bin for the LRG and control samples, respectively. }
\label{fig:control}
\end{figure}

Following the radio galaxy matching procedure of Section \ref{matching}, we searched for double-lobed radio galaxies (FR I/II) around the optical LRG positions.  The matches discovered in this way, as well as the false positives found with the control sample, are plotted for the photo-z LRG sample in Figure \ref{fig:control}.   Shown are the number of matches versus lobe opening angle ($\theta$) for progressively larger lobe separations.  The excess of LRG matches over random matches indicates the number of true matches found.  As expected, increasing the allowable lobe separation causes the contamination by random matches to go up.  Although Figure \ref{fig:control} shows that the excess over random matches declines very quickly above lobe separations of 60", we find that allowing matches out to 350" adds signal and increases the completeness from 65\% ($\pm 6\%$) at 120" to 96\% ($\pm 10\%$).  The excess drops to zero above lobe separations of 350", so we will only include as matches FR I/IIs with lobe separations less than this value for the percentages quoted below.


Small values of $\theta$ (opening angle) indicate bent double-lobed sources.  Bent double radio galaxies are not unexpected and are hypothesized to result from a number of situations, from deflection due to interactions between the jets and dense gas clouds in the narrow line region \citep{man98} to bending due to the ram pressure of the intracluster medium (e.g., Vallee, Wilson, \& Bridle 1981; Sakelliou \& Merrifield 2000).  However, small opening angles are also caused by nearby but unrelated pairs of radio sources simply posing as lobes, and this possibility increases in likelihood for progressively smaller opening angles.  The excess of double-lobed candidates with small opening angles ($\theta$ $< 90^{\circ}$) seen in Figure \ref{fig:control} is therefore likely to be due to the clustering of other galaxies with the LRGs.  The significant excess of matches with small opening angles seen in the left-most panel, and the fact that the excess is fairly constant even to below $50^{\circ}$, where real double-lobed sources are rare, are indications that we are detecting clustering in the environments of the LRGs.   This is consistent with the notion that LRGs are highly clustered, tending to reside at the centers of galaxy groups and clusters (Zehavi et al. 2005; Zheng et al. 2008).  Moreover, the effect here has been downplayed by our technique of excluding radio sources with optical counterparts, as many of the cluster galaxies will presumably have optical emission.  

\begin{figure*}[ht]
\centering
\includegraphics[scale=0.8]{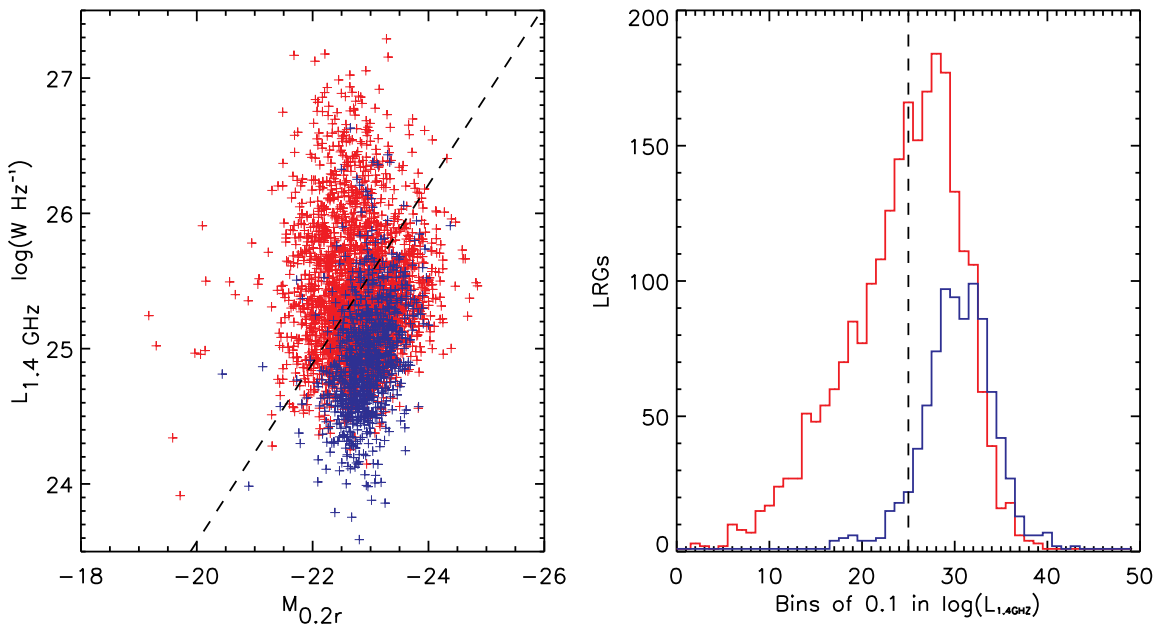}
\caption{Left panel: Rest frame 1.4 GHz radio luminosity in W Hz$^{-1}$ vs. r-band absolute magnitude, k+e corrected to 0.2.  The blue points are FR I/II candidates from the spec-z LRG sample, and the red points are candidates from the photo-z LRG sample.  The solid line is from \citet{led96} to divide FR I/II mophologies, with sources falling above the line defined as FR IIs and below the line as FR Is. Right panel: a histogram constructed from the same plot by taking bins of 0.1 in log($L_{1.4GHz}$) parallel to the divide.  $M_r$ increases to the right as with the left panel.}
\label{fig:FRdivide}
\end{figure*}

Our initial matching for the photo-z LRGs produced 542,557 candidate double-lobed radio galaxy matches.  We then calculated the excess over random matches and restrict the lobe opening angle to $\theta$ $> 140^{\circ}$ to minimize the clustering effect.  This procedure leaves 4376 LRGs associated with FR I/II double-lobed morphologies, or 0.57\% ($\pm$0.02) of the LRG sample.  The contamination rate is 83.1\%, indicating that although we can constrain the overall FR I/II percentage quite well, there is a significant background signal.  For the spec-z LRG sample, 1.92\% ($\pm$0.09) are associated with FR I/II morphologies, with a contamination rate of 58.1\%. The higher number of FR I/II candidates is most likely primarily due to the lower mean redshift of the sample, since the FIRST catalog is flux-limited.


Unfortunately, FIRST does not have the resolution or sensitivity to further classify the majority of these radio galaxies into the typical FR I and FR II classes \citep{fan74} based on morphology.  We therefore utilize the result of \citet{led96} who found that the FR I/II break is a strong function of the host galaxy's optical luminosity.  We create a sample of ``most likely" double-lobed LRGs by selecting those with lobe separation $<$ 1' and opening angle $\theta > 150^{\circ}$.  The resulting sample consists of 2411 double-lobed LRGs from the photo-z LRG sample (96.5\% of which are real based on the control sample), and 880 from the spec-z LRG sample (of which 99\% are real).  We overplot the Ledlow \& Owen (1996) dividing line on our ``most likely" sample in the left panel of Figure \ref{fig:FRdivide}.   The blue points come from the spec-z LRG sample and the red points come from the photo-z LRG sample.  Here we k+e correct the r-band magnitude to $z=0.2$ for consistency with Ledlow \& Owen, and we assume M$_r$ $=$ M$_{24.5}$, the isophotal magnitude to 24.5 mag arcsec$^{-2}$ as they use.  We have derived the flux densities from the sum of the NVSS components, since FIRST's higher resolution may resolve out more diffuse components.  To ensure that we only include flux from the associated radio lobes and core (if present) and not from nearby, unrelated radio sources, we follow \citet{sad07} and reduce the NVSS flux by the flux ratio of the associated to unassociated FIRST sources in cases where NVSS cannot resolve the individual lobes.   It is true that, in some cases, FIRST has resolved a radio galaxy into more than 3 components, and in these cases the ``corrected" radio luminosities reported here are really lower limits.  In the $< 1\%$ of the cases where the NVSS flux is zero, we use the values from the FIRST catalog.

We find that photo-z LRGs in the ``most likely" double-lobed radio galaxy sample fall into both morphological classes.  The majority (56\%) of the double-lobed galaxies fall into the lower luminosity FR I classification, while 44\% appear to be FR IIs.  This is a larger percentage of FR IIs than that found for a similar sample by Sadler et al. (2007 - their Figure 14).  This is mainly due to the fact that \citet{sad07} use different matching criteria, resulting in a vast majority (81\%) of their radio galaxies having just \textit{one} associated FIRST component.  They use the luminosity of this single component to place the galaxy on this plot.

For the spec-z LRGs, Figure \ref{fig:FRdivide} shows that most of the ``most likely" sample (93\%) fall into the FR I classification.  There are two effects that cause this apparent discrepancy between the spec-z and photo-z LRG samples.  The first is that the detection of FR Is in the photo-z LRG sample is suppressed because these objects are typically fainter in the radio than FR IIs and further away than the spec-z LRGs on average.  The strong optical luminosity dependence of the Ledlow \& Owen relation is a contributing factor as well since the the photo-z LRGs have more optically-fainter objects, where the threshold to be classified as an FR II is lower.   When we impose a luminosity threshold of M$_{0.2r}$ $<$ -22.5 and compute the percentage of FR I and FR II galaxies separately (as a percentage of the total LRG sample), we find that the photo-z LRGs have 0.06\% FR IIs while the spec-z LRGs have a comparable 0.07\%.  For FR Is, the photo-z LRGs have 0.15\% while the spec-z LRGs have 1.11\%, or over seven times more.  This remaining discrepancy is most likely due to redshift and luminosity-dependent selection effects.  


The right panel of Figure \ref{fig:FRdivide} shows the distribution of radio galaxies that make up the left panel.  Bins are log($L_{radio}) = 0.1$ in width and symmetric about the Ledlow \& Owen dividing line.  As the histograms in the right panel show, the distribution of galaxies shows no sign of any bimodality.  Clearly, therefore, the exact quantitative results obtained using this method should be approached with healthy skepticism;  A small shift in the dividing line would lead to a significant change in the calculated percentages of FR I's versus FR II's.  Moreover, the line itself remains an imperfect discriminator, and thus there is a large uncertainty for the classifications drawn from this luminosity-based technique.    






\begin{table}[ht]
\centering
  \caption{Median Properties of the LRG Samples}
  \begin{tabular} {llcc}
       \hline\hline
       Sample & L$_{1.4 GHz}$ & z & M$_i$ \\
       & (ergs s$^{-1}$ Hz$^{-1}$)  & & (mag) \\
       \hline
       Photo-z LRGs & &\\
       ALL & 4.11 ($\pm$0.03) $\times 10^{29}$ & 0.53 & -22.9\\
       FR I/IIs & 219 ($\pm$7) $\times 10^{29}$ & 0.56 & -23.4\\
       \hline
       Spec-z LRGs & &\\
       ALL & 3.85 ($\pm$0.05) $\times 10^{29}$ & 0.35 & -23.3\\
       FR I/IIs & 92 ($\pm$4) $\times 10^{29}$ & 0.34 & -23.5\\
       \hline
  \end{tabular}
  \label{tab:stack}
\end{table}

\subsubsection{Dependence on Radio Luminosity}
\label{LRGrates}

We next investigate how the presence of a double-lobed morphology depends on core radio luminosity.  We do this by comparing the stacked luminosity of the ``most-likely" FR I/II LRGs with the stacked luminosity of the LRG samples as a whole.  The results are presented in Table \ref{tab:stack} for both the photo-z and spec-z LRG samples.  We leave the samples separate as they have significantly different sample redshifts.  With only a small difference in median redshift, the cores of the LRGs that are also double-lobed radio galaxies are 20-50 times more luminous in the radio than those that are not doubles.  This agrees qualitatively with the findings of \citet{deV06}, who report that quasars that are also FR IIs have a much higher percentage of FIRST-detected cores than the optical quasar population in general.  The double-lobed LRGs are also more luminous in the optical, as expected from their higher core radio luminosities (see, for e.g., Auriemma et al. 1977), with a median i-band magnitude 0.2-0.5 mags brighter than the sample median.  The photo-z LRGs show a 0.4 mag difference in median M$_r$, which agrees with that found by \citet{sad07} for intermediate-redshift LRGs in the 2SLAQ sample. 

\begin{figure}[ht]
\centering
\includegraphics[scale=0.45]{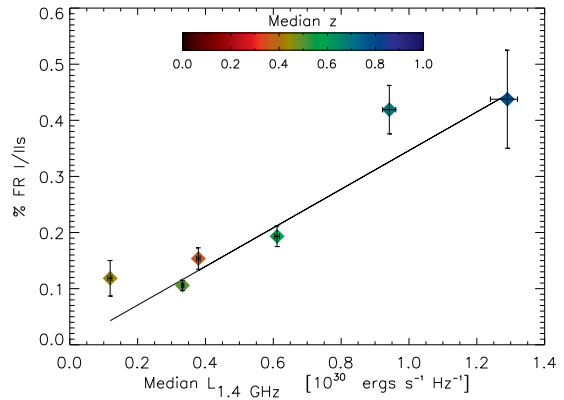}
\caption{FR I/II \% vs. median radio luminosity for both photo-z and spec-z LRGs.  The photo-z LRG sample was binned by redshift, and the color bar shows the median redshift associated with each point.  The third point from the left (at $\sim$0.4 $\times$ $10^{30}$ ergs s$^{-1}$ Hz$^{-1}$) represents the spec-z LRG sample, binned altogether to improve statistics.}
\label{fig:LRGrates}
\end{figure}

We then split the LRG sample into 5 redshift bins and apply the radio galaxy matching technique to calculate percentages for each bin.  To counter the flux-limited nature of the FIRST catalog, we impose a luminosity threshold calculated by assuming a 1 mJy source at $z = 1$.  The results are plotted in Figure \ref{fig:LRGrates}, where the third point from the left is the result of stacking the spec-z LRG sample altogether.  The percentage of LRGs that host double-lobed radio galaxies increases significantly with increasing median radio luminosity.  The best-fitting line to the weighted data points yields FR I/II $\% = (0.34 \pm 0.04) L_{radio}$ - $(0.00 \pm 0.02)$, where $L_{radio}$ is in units of 10$^{30}$ ergs s$^{-1}$ Hz$^{-1}$.  Since the LRG sample was sorted by redshift, it is possible that a redshift-dependence or evolutionary effect is contributing to the steepness of this trend.  The radio detection rate increases with redshift in general since the massive galaxies that host radio sources constitute a larger fraction at higher redshift \citep{sad07}.  However, the spec-z LRG data point argues against this theory, since the data would no longer show such an obvious trend if plotted versus redshift.  Therefore, although we cannot completely rule out the possibility that a redshift dependence is contributing to the trend seen here, it is likely not the dominant factor.

What is causing the trend? Core-detected galaxies lie in the bright-end tail of the distribution of core radio luminosities.  If we can assume that the shape of this distribution is fairly uniform over our redshift range, such that the entire distribution simply shifts upward in radio luminosity, then it's not hard to imagine that as the median radio luminosity increases, more of the LRGs achieve the core luminosity necessary to launch powerful FR I/II jets.  However, there is a strong selection bias is at play here.  While we stack to get the core luminosity, the detection of lobes is still limited by the 1 mJy flux limit of the FIRST survey.  For a given distribution of core luminosities, and assuming a fixed average core-to-lobe ratio, shifting the distribution to higher core luminosities would mean that more lobes would become \textit{detectable}.  This selection bias means that we cannot conclude whether there is a threshold core luminosity required to launch jets in general.  We can say what a typical core-to-lobe ratio is for this class of objects, since we are sensitive to very faint core luminosities.  For the faintest LRGs to show up as FR I/IIs, the lobes would have to be over 100 times brighter than the cores.  For half to show up, they would have to be around 40 times brighter.  Using the median luminosity of our stacked FR I/II sample, the lobes of these FR I/IIs are typically only about 5 times brighter (core-to-lobe ratio R $\simeq$ 0.2).  These values imply a distribution of core-to-lobe ratios similar to others in the literature for samples of FR I/II radio galaxies (e.g., Zirbel \& Baum 1995).

\subsubsection{Comparison to Quasars}

\begin{figure}[ht]
\centering
\includegraphics[scale=0.45]{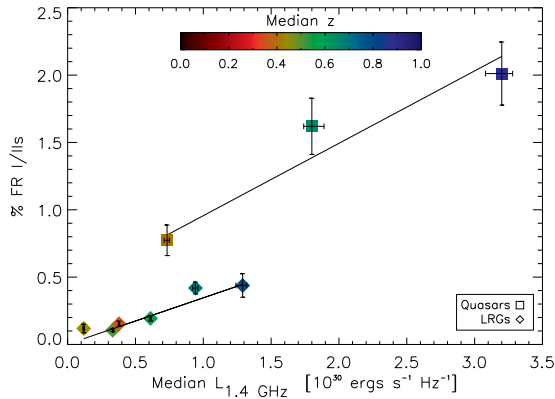}
\caption{FR I/II \% vs. median radio luminosity as in Figure \ref{fig:LRGrates}, but with rates for quasars with $z < 1$ overplotted as squares.  }
\label{fig:ALLrates}
\end{figure}

As a comparison sample, we took the SDSS DR5 quasar catalog \citep{sch07} and split the quasars below $z = 1$ into three redshift bins.  (We limited the redshifts to $z < 1$ to match the range of the LRG sample and to keep the luminosity threshold low.)  The resulting percentages of double-lobed radio galaxies are plotted over the LRGs in Figure \ref{fig:ALLrates}.  For a given median radio luminosity, quasars have a higher incidence of double-lobed radio galaxies than LRGs, although the rates are within a factor of 2-3.  They also show a slightly stronger dependence on median luminosity, with FR I/II $\% = (0.5 \pm 0.1) L_{radio}$ + $(0.4 \pm 0.2)$, where again the units of $L_{radio}$ are $10^{30}$ ergs s$^{-1}$ Hz$^{-1}$.  The higher rates seen for the quasars may be a selection effect; by targeting quasars, we are observing objects whose AGN are (by definition) in an active phase.  Even though these quasars were selected via optical activity, roughly 10\% of quasars are also radio-loud, and even those that are ``radio-quiet" are not radio silent; most have some level of radio emission.  Thus, these results may reflect a difference in the duty cycles for the radio AGN activity of quasars and LRGs.  


We conclude by noting that when we include higher-redshift quasars, the trend seen in Figure \ref{fig:ALLrates} does not continue, but turns over. (The exact redshift of the turnover depends on the choice of luminosity threshold.)  This turnover is simply an artifact of surface brightness dimming, which goes as $(1 + z)^{4}$.  This steep redshift dependence biases against extended sources at higher redshift, causing them to no longer make it into the catalog. 




\section{DISCUSSION}
\label{Discussion}

It has been shown previously (Paper III) that LRGs from the SDSS spectroscopic sample host low-luminosity radio AGN.  Taking this sample to be representative of the larger, photo-z LRG sample introduced in this work, we can assume that the median star formation rate is $\sim$ 1 M$_{\sun}$ yr$^{-1}$.   Using the relation between SFR and radio luminosity derived by Paper III, and taking the median redshift of the photo-z LRG sample, we therefore estimate that the median radio flux density attributable to star formation is 0.5 $\mu$Jy, or 1\% of the total median flux density of these LRGs.  If one believes the 4000-\AA\ derived star formation rates referenced in Paper III, which extend beyond 20 M$_{\sun}$ yr$^{-1}$ for these red and dead galaxies, then that still ensures that for over 90\% of the LRGs, the contribution to the radio luminosity from star formation will be an order of magnitude smaller than the contribution from AGN activity.  


The radio emission is therefore evidence that nuclear accretion is occurring in these early-type galaxies despite the lack of optical emission lines (Paper III).  The absence of emission lines is likely due to dilution of the lines by the stellar light from the host galaxy.   The supermassive black holes in these massive red galaxies have presumably used up much of their gaseous reservoirs and are now accreting at a lower rate.  If the accretion onto the black hole has transitioned to a radiatively inefficient accretion flow, as suggested by work in the X-rays \citep{bra05}, then it is likely that the radiative signature has weakened beyond detectability.

Our discovery that low-luminosity radio AGN are characteristic of the LRG population also agrees with recent models of galaxy formation (Croton et al. 2006; Cattaneo \& Teyssier 2007).  In such models, radio-mode AGN feedback is thought to aid massive red galaxies in suppressing further star-formation.  The discovery of radio AGN in intermediate redshift LRGs in general, not just the small percent detected in FIRST, provides a means for this to occur.  

We find that the radio luminosity of the AGN shows a correlation with the optical luminosity of the LRG.  
This correlation is not simply the result of radiation being reprocessed at another wavelength (for example, optical photons absorbed and re-emitted in the infrared by dust grains).  While the radio emission originates from the active nucleus, the optical emission comes mainly from the stellar component of the host galaxy.  Even when the optical emission is dominated by the nuclear component, the emission regions are still physically distinct.  In that case, the correlation indicates a link between the fueling of relativistic jets and accretion onto a black hole which produces the optical continuum (Serjeant et al. 1998; Kukula et al. 1998; Willott et al. 2000).  

A radio-optical correlation has been determined for many different galaxy classes, although work in the past has always relied on samples with individual radio detections.  The current work is different in this regard, and the sample so faint that no comparisons are really applicable.  The closest comparison available might be that of \citet{whi07}, who looked at the radio-optical correlation of quasars in Paper I using the same technique.  The relation they derived was $L_{radio} \propto L_{opt}^{0.85}$, and they found the slope of the power law to be independent of redshift interval.  We would not necessarily expect an equivalent result, as the optical emission from the quasars is more likely to be dominated by the central AGN.  

After accounting for the radio-optical correlation, we find that the radio AGN are evolving, implying a tailing off of accretion activity.  Our most conservative estimate, which comes from simultaneously fitting the luminosity and redshift dependencies, infers fairly significant evolution of the form $L_{1.4GHz}$ $\sim$ (1+z)$^{\alpha}$ with $\alpha = 3.15 \pm 0.07$.  Here we cannot say whether the physical mechanism behind the evolution is density or luminosity evolution; we can only conclude that evolution has occurred.  

This result conflicts with \citet{cle04}, who found no evidence for evolution with redshift of low-luminosity radio sources, but agrees with the results of \citet{bro01}. The implied evolution is also comparable to favored estimates for the evolution of the SFR density with redshift (Hopkins 2004).  It is somewhat stronger than the evolution measured by both \citet{sad07} and \citet{don09} for radio LRGs in particular.  \citet{sad07} show that the low-luminosity ($L_{1.4GHz} < 10^{32}$ ergs s$^{-1}$ Hz$^{-1}$) radio galaxies associated with 2SLAQ LRGs are well-fit by pure luminosity evolution of the form $(1 + z)^{k}$ where $k = 2.0 \pm 0.3$ (overplotted in Figure \ref{fig:rlumvsMi_slopes} as a dotted line).  \citet{don09} measure the evolution using the MegaZ-LRG catalog, concluding that neither pure luminosity nor pure density evolution provide a good fit, but that the comoving number density of radio AGN with luminosities less than 10$^{32}$ ergs s$^{-1}$ Hz$^{-1}$ increases by a factor of $\sim$ 1.5 between z $=$ 0.14 and z $=$ 0.55.  

This work constitutes one of the first looks at the evolutionary history of a typical low-luminosity radio AGN residing in an LRG.  The LRG population is dominated by these low-luminosity AGN, with median flux densities placing them well into the submillijansky regime.  The associated median radio luminosities for these intermediate redshift LRGs ($L_{1.4 GHz}$ $<$ 10$^{30}$ ergs s$^{-1}$ Hz$^{-1}$) are comparable with those of the bright end of the local low-luminosity AGN distribution.  The median luminosity is at least an order of magnitude fainter than LRGs typically detectable in FIRST.  Using the \citet{bes06} conversion between radio luminosity and mechanical energy input, and extrapolating to lower redshift, the strong evolution we detect suggests that the total energy input from AGN heating by massive early-type galaxies was approximately 50\% higher at z$\sim$0.5.  This is in rough agreement with the evolution predicted by \citet{sad07} by extrapolating the luminosity evolution of FIRST-detected low-power radio galaxies. 





As an added complexity to the overall evolution, Section \ref{downsizing} demonstrates that the slope of the L$_{radio}$-L$_{optical}$ relation varies with redshift, increasing down to $z \simeq 0.5$, then perhaps approaching a constant value in the local universe (the error bars are too large to draw a firm conclusion here).  An increase in slope with decreasing redshift would imply that lower absolute magnitude galaxies are experiencing stronger evolution in radio luminosity than higher absolute magnitude galaxies.  This would be perhaps indicative of AGN cosmic downsizing, in which lower-luminosity AGN peak in their comoving space density at lower redshift than higher-luminosity AGN.   This effect has been seen, for example, in X-ray studies tracing the AGN number density as a function of luminosity (Cowie et al. 2003; Hasinger et al. 2005), and by comparing optically-derived normalized accretion rates as a function of host galaxy mass \citep{kri07}.  Recent work by \citet{rev08} is also suggestive of downsizing: they find luminosity-dependent evolution of radio power by stacking emission-line AGN from the SDSS as a function of host luminosity.  While our error bars are too large to constrain any evolution of L$_{radio}$-L$_{optical}$ below about $z = 0.4$, at higher redshifts we do see a monotonic increase in evolution with decreasing optical luminosity.  This may be evidence of downsizing, and if so, it extends the work of \citet{rev08} from lookback times of 2 Gyr to lookback times of over 6 Gyr.    


\section{CONCLUSIONS}
\label{conclusions}

We have used the FIRST survey to investigate the radio properties of over 65,000 low redshift and 760,000 intermediate redshift optically-selected LRGs.  As the sample is largely ``radio-quiet", we used a median-stacking technique to achieve the required sensitivity and study trends in the AGN-driven radio luminosity of the sample as a whole.  We also matched the LRG positions with the FIRST survey catalog to study the sub-population associated with radio-loud double-lobed morphologies.  Our main conclusions are summarized here.  

\begin{itemize}
\item By median stacking, we find that the spec-z (low-z) LRGs have typical 1.4 GHz flux densities in the 100s of $\mu$Jy, while the photo-z (intermediate-z) LRGs have typical values in the 10s of $\mu$Jy.  For their respective redshift ranges, these flux densities correspond to radio sources in the range 10$^{29}$ $<$  $L_{1.4GHz}$ $<$ 10$^{30}$ (ergs s$^{-1}$ Hz$^{-1}$).  The contribution to the radio power from star formation is over an order of magnitude lower for 90\% of the sources, implying that LRGs host low-luminosity radio AGN.  

\item We find that the median radio luminosity scales with the optical luminosity as $L_{1.4GHz} \propto L_{opt}^{\beta}$, where $\beta$ depends on the redshift being probed.  The value of $\beta$ appears to decrease between redshifts of $z = 0.45$ and $z = 0.75$, suggesting that lower-luminosity LRGs are experiencing stronger evolution in their radio power than higher-luminosity LRGs over this redshift range.  This may be a signature of AGN downsizing.  Below $z \sim 0.4$, the data are consistent with $\beta$ $=$ constant.

\item We present evidence that the radio AGN in LRGs undergo fairly significant cosmic evolution over the range 0.2 $<$ z $<$ 0.7.  A simultaneous least-squares fit to the redshift and luminosity dependencies yields redshift evolution of the form $L_{1.4GHz} \propto (1+z)^{3.15 \pm 0.07}$.  There could be a range of physical explanations behind this evolution, from all sources getting brighter, to longer duty cycles for constant-luminosity sources.  Either way, the total AGN heating due to these massive ellipticals would have been considerably larger at higher redshift.  

\item Additionally, a small percentage of the LRG sample host double-lobed (FR I or FR II) radio galaxies.  Searching for matches with the FIRST catalog, we find that 0.57\% ($\pm$0.02) of the photo-z LRGs are radio doubles, while the incidence is 1.92\% ($\pm$0.02) for the lower redshift spec-z LRGs.  We find an excess of double-lobe candidate matches at small values of the lobe opening angle.  This is likely an artificial signal due to the excess clustering in the environments of LRGs.  

\item We separate the most likely double-lobe candidates into FR I/IIs based on luminosity.  We find that the intermediate-redshift photo-z LRGs fall into both FR I and FR II classes, while the vast majority of the low-redshift spec-z LRGs are FR Is.  This difference is likely due to redshift and luminosity selection effects.  Neither sample of galaxies show any sign of bimodality.  

\item The cores of LRGs associated with FR I/IIs are 20-50 times more powerful than the general LRG sample, and the percentage of FR I/IIs increases with increasing median stacked radio luminosity.  These trends are likely due to a strong selection bias where, for a given distribution of core luminosities, and assuming a fixed average core-to-lobe ratio, increasing core luminosity means more lobes exceed FIRST's 1 mJy flux limit.  Our results imply that LRGs have a median core-to-lobe ratio of $\sim$0.2.  

\end{itemize}

In conclusion, we have taken advantage of the overlapping coverage of FIRST and SDSS to learn more about the radio emission from AGN in the LRG population, high and low-luminosity alike.  As more sensitive telescopes come online and deeper, wide-field surveys are produced, researching the general properties of submillijansky populations such as this will become the norm.  At the same time, the stacking procedure used here will continue to push the boundaries of what is truly radio-silent.

\acknowledgements
The authors would like to thank the anonymous referee for comments which resulted in the significant improvement of this paper.  We also wish to thank David Schlegel and Antonio Cuesta Vazquez for kindly providing the photometric catalog of LRGs on which this paper is based.  JAH acknowledges support of a UC Davis Graduate Block Grant Fellowship, NRAO Grant GSSP08-0034, and Grant HST-GO-10412.03-A from the Space Telescope.  GRZ also acknowledges NRAO Grant GSSP08-0034.  RHB acknowledge the support of the National Science Foundation under grant AST 00-98355.  RLW acknowledges the support of the Space Telescope Science Institute, which is operated by the Association of Universities for Research in Astronomy under NASA contract NAS5-26555.  The work by RHB was partly performed under the auspices of the U.S. Department of Energy by Lawrence Livermore National Laboratory under Contract DE-AC52-07NA27344.  

The National Radio Astronomy Observatory is a facility of the National Science Foundation under cooperative agreement by Associated Universities, Inc.

\end{document}